\documentclass[preprint2]{aastex_pvd}

\usepackage{epsf,apjfonts}

\gdef\msun{M$_{\odot}$}
\gdef\Ja{$J_1$}
\gdef\Jb{$J_2$}
\gdef\Jc{$J_3$}
\gdef\Ha{$H_1$}
\gdef\Hb{$H_2$}
\begin{document}

\title{The NEWFIRM Medium-Band Survey: Filter Definitions and First
Results}

\author{Pieter G.\ van Dokkum\altaffilmark{1,2}, Ivo Labb\'e\altaffilmark{3},
Danilo Marchesini\altaffilmark{1}, Ryan Quadri\altaffilmark{4},
Gabriel Brammer\altaffilmark{1}, Katherine E.\ Whitaker\altaffilmark{1},
Mariska Kriek\altaffilmark{5}, Marijn Franx\altaffilmark{4},
Gregory Rudnick\altaffilmark{6}, Garth Illingworth\altaffilmark{7},
Kyoung-Soo Lee\altaffilmark{1}, Adam Muzzin\altaffilmark{1}}


\begin{abstract}
Deep near-infrared imaging surveys allow us to select and study distant
galaxies in the rest-frame optical, and have transformed our understanding
of the early Universe. As the vast majority of $K$- or IRAC-selected
galaxies is too faint for spectroscopy, the interpretation of these surveys
relies almost exclusively
on photometric redshifts determined from fitting templates to the
broad-band photometry. The best-achieved accuracy of these redshifts
$\Delta z / (1+z) \gtrsim 0.06$ at $z>1.5$, which is sufficient for determining
the broad characteristics of the galaxy population but not for measuring
accurate rest-frame colors, stellar population
parameters, or the local galaxy density. We have started a
near-infrared imaging survey with the NEWFIRM camera on the Kitt Peak 4m
telescope to greatly improve the accuracy
of photometric redshifts in the range $1.5\lesssim z \lesssim 3.5$.
The survey uses five medium-bandwidth filters,
which provide crude ``spectra'' over the wavelength range $1-1.8\,\mu$m
for all objects in the $27\farcm 6 \times 27\farcm 6$ NEWFIRM field.
In this first paper, we illustrate the technique by showing
medium band NEWFIRM
photometry of several galaxies at $1.7<z<2.7$ from the near-infrared
spectroscopic sample of Kriek et al.\ (2008). The filters unambiguously
pinpoint the location of the redshifted Balmer break in these galaxies,
enabling very accurate redshift measurements.
The full survey will provide similar
data for $\sim 8000$ faint $K$-selected galaxies at $z>1.5$
in the COSMOS and
AEGIS fields. The filter set also enables efficient
selection of exotic objects such as high redshift quasars,
galaxies dominated by emission lines, and very cool brown dwarfs;
we show that late T and candidate ``Y''
dwarfs could be identified using only two of the
filters.

\end{abstract}


\altaffiltext{1}{Department of Astronomy, Yale University, New Haven, CT 06520-8101.}
\altaffiltext{2}{Visiting Astronomer, Kitt Peak National Observatory, National Optical Astronomy Observatory, which is operated by the Association of Universities for Research in Astronomy (AURA) under cooperative agreement with the National Science Foundation.}
\altaffiltext{3}{Carnegie Observatories, Pasadena, CA 91101.}
\altaffiltext{4}{Sterrewacht Leiden, Leiden University, NL-2300 RA Leiden, Netherlands.}
\altaffiltext{5}{Department of Astrophysical Sciences, Princeton University, Princeton, NJ
08544.}
\altaffiltext{6}{Department of Physics and Astronomy, University of Kansas, Lawrence, KS 66045.}
\altaffiltext{7}{UCO/Lick Observatory, University of California, Santa Cruz, CA 95064.}

\section{Introduction}

It has become clear that the Universe at $1.5<z<3.5$ saw a much greater
diversity of galaxies than the Universe today. This epoch is often
characterized as one of rapid change, as
many galaxies were experiencing 
strong and presumably short-lived star formation ({Steidel} {et~al.} 1996; {Blain} {et~al.} 2002),
significant merging activity, and rapid black
hole growth (e.g, {Daddi} {et~al.} 2007). At the same time
a substantial population of
quiescent galaxies already existed
(e.g., {Kriek} {et~al.} 2006). These galaxies have spectra characterized by
strong Balmer or 4000\,\AA\ breaks and no detected H$\alpha$
emission. They also have very compact morphologies, which implies they
must undergo significant subsequent evolution
(e.g., {Trujillo} {et~al.} 2006; {Cimatti} {et~al.} 2008; {van Dokkum} {et~al.} 2008).

Given the diversity
and rapid changes in the galaxy population at this epoch, it is important
to secure and study large, uniformly selected samples of galaxies
at $1.5<z<3.5$ in a homogeneous way.
Unfortunately it is difficult to obtain such samples,
as familiar rest-frame
optical spectral features are shifted into the near-infrared.
As a consequence,
most studies of high redshift galaxies have focused on
blue star forming galaxies, as they are relatively bright
at observed optical (rest-frame ultra-violet) wavelengths
(e.g., {Steidel} {et~al.} 1996, 1999). Although this approach
is very efficient, it misses galaxies that are relatively red:
as shown in {van Dokkum} {et~al.} (2006) the majority of galaxies with
masses $>10^{11}$\,\msun\ at $2<z<3$ would not be selected by
traditional Lyman break criteria. Studying samples that are
complete to a rest-frame optical limit, or to a stellar mass
limit, implies a compromise: one either
works with small, bright samples for which it is possible
to obtain spectra (Cimatti et al.\ 2002; Kriek et al.\ 2008),
or one relies on photometric redshifts derived from broad-band
photometry
(e.g., {Dickinson} {et~al.} 2003; {Fontana} {et~al.} 2006, and many other studies).
These photometric redshifts
are generally assumed
to be sufficiently accurate for determining broad characteristics
of the galaxy population, such as the luminosity function. However, redshift
errors can lead to biases (see, e.g., Marchesini
et al.\ 2007, Reddy et al.\ 2008), and these redshifts cannot
be used to measure accurate rest-frame colors, stellar population
parameters, or the local galaxy density. Furthermore, their
random errors typically have a non-Gaussian distribution
(leading to so-called ``catastrophic failures'') and they
can have significant systematic uncertainties
(see, e.g., Brammer et al.\ 2008, and references therein).

Inspired by the successful COMBO-17 optical medium-band
imaging survey at redshifts $0<z<1$ ({Wolf} {et~al.} 2003),
we are undertaking a project which will provide a sample of $K$-selected
galaxies with accurate redshifts in the range $1.5<z<3.5$ that is several
orders of magnitude larger than what is available today.
We designed and manufactured a set of five medium-bandwidth near-infrared
(near-IR) filters, which provide ``spectra'' with a resolution
of $R\sim 10$ from $1-1.8\,\mu$m. The filters are designed to isolate
the location of the redshifted Balmer- or 4000\,\AA-break for
galaxies at $1.5<z<3.5$. A set of these filters was manufactured for
the NEWFIRM camera ({Probst} {et~al.} 2004) on the Kitt Peak 4m, and the
NEWFIRM Medium Band Survey (NMBS; an NOAO Survey Program) began
in March 2008. This paper describes the characteristics of the filters,
outlines the survey strategy,
and shows results from a short pilot program that we executed in
the Spring of 2008. The survey will be described more extensively
in a forthcoming
paper (K.\ Whitaker et al., in preparation).

\section{Filter Characteristics}

\begin{figure*}[t]
\epsfxsize=11.2cm
\epsffile[-35 252 429 665]{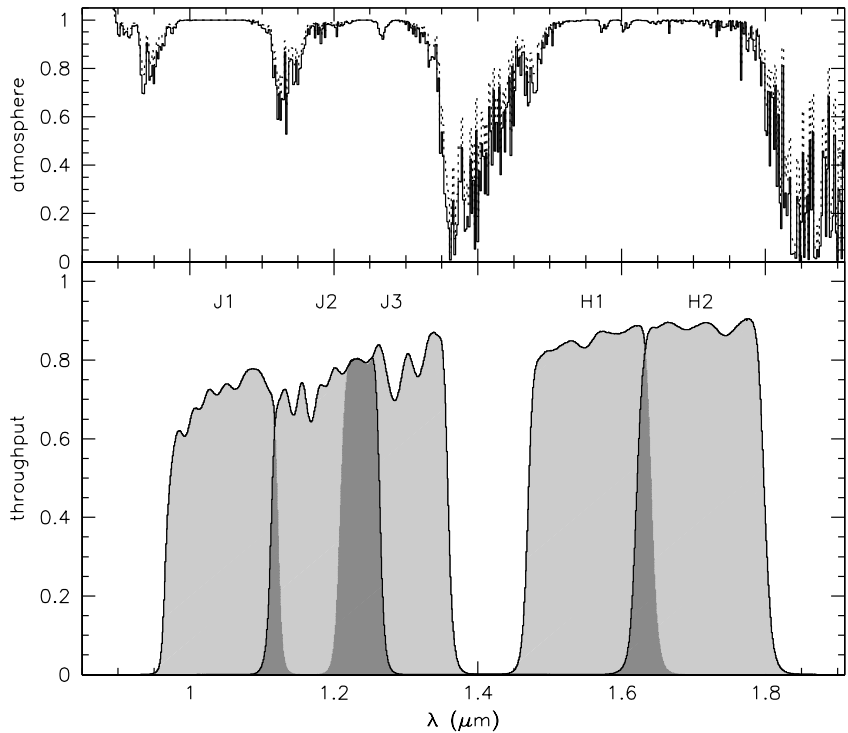}
\caption{\small Medium-bandwidth filters designed
for NEWFIRM and used in the NMBS. The throughput of the
filters ranges from $\approx 70$\,\% for \Ja\ to
$\approx 90$\,\% for \Hb\ (excluding effects of
the atmosphere). The top
panel shows the atmospheric transmission spectrum, for two different
water columns: the broken line is for a column of 1.6\,mm and the
solid line is for 3.0\,mm. 
\label{filters.plot}}
\end{figure*}

The filters are shown in Fig.\ \ref{filters.plot}. The $J$ band is
split in three filters \Ja, \Jb, and \Jc, and the $H$ band
is split in two filters \Ha\ and \Hb. Each filter consists of
two physically separate
components: a transmission filter and a blocking filter, which both
need to be mounted in the two filter wheels of NEWFIRM. The blocking
filters are needed because of the long-wavelength sensitivity of
NEWFIRM's InSb arrays, and are the cause of the ``roll-off'' in
sensitivity toward shorter wavelengths and the wiggles in
the transmission curves of (particularly) \Jb\ and \Jc.
The \Ja\
filter has its own blocking filter, \Jb\ and \Jc\ share a blocking
filter, and \Ha\ and \Hb\ share a blocking filter. Mounting all
five filters in NEWFIRM therefore requires eight filter slots.

Central wavelengths and 50\,\% cut-on and cut-off wavelengths
for the five filters are listed in Table 1.
The \Ja\ filter is somewhat redder and broader
than the $Y$ filter used by,
e.g., the UKIRT Infrared Deep Sky Survey, whose
cut-on and cut-off wavelengths are $0.97\,\mu$m and $1.07\,\mu$m
respectively. The \Jc\ filter is narrower and redder than the
$J_s$ (``$J$-short'') filter used in, e.g., HAWK-I on the
Very Large Telescope. The wavelength range
covered by the \Jb\ filter contains an atmospheric
H$_2$O absorption feature, and the red edge of
the \Jc\ filter pushes slightly into the H$_2$O band between
the $J$ and $H$ windows. The atmosphere leads to a decrease
in throughput of up to $\approx 10$\,\% in the \Jb\ and \Jc\
filters. Figure \ref{filters.plot}
and Table 1 demonstrate the effects of varying the
H$_2$O column between 1.6\,mm and 3.0\,mm; for all filters the
variations in throughput due to variations in the water column
are $\lesssim 2$\,\%.

\begin{table}[h]
\small
\caption{Basic Filter Data}
\begin{tabular}{lccccc}
\hline
\hline
 & \Ja\ & \Jb\ & \Jc\ & \Ha\ & \Hb\ \\
\hline
$\lambda_{\rm cen}$\tablenotemark{a} & 1.047 & 1.195 & 1.279 & 1.560 & 1.708 \\
$\lambda_{\rm blue}$\tablenotemark{b} & 0.969 & 1.115 & 1.209 & 1.474 & 1.621 \\
$\lambda_{\rm red}$\tablenotemark{b} & 1.120 & 1.263 & 1.350 & 1.641 & 1.796 \\
$T_{1.6\,{\rm mm}}$\tablenotemark{c} & 0.98 & 0.94 & 0.92 & 0.97 & 0.97 \\
$T_{3.0\,{\rm mm}}$\tablenotemark{c} & 0.97 & 0.92 & 0.90 & 0.96 & 0.96 \\
$\Delta$\,AB\tablenotemark{d} & 0.65 & 0.84 & 0.96 & 1.29 & 1.42 \\
\hline
\end{tabular}

\tablenotetext{a}{Includes atmosphere, with 3.0\,mm water column.}
\tablenotetext{b}{50\,\% cuton/cutoff wavelengths; includes atmosphere.}
\tablenotetext{c}{Transmission of the atmosphere for different water columns.}
\tablenotetext{d}{Synthetic AB magnitudes of Vega.}
\end{table}

Transformations from the Vega to the AB system were calculated by
integrating the Vega spectrum, using filter curves that include
atmospheric absorption. The uncertainties in the AB offsets listed
in Table 1 are $\approx 0.02$, and are dominated by uncertainties
in the absolute calibration of Vega.

\section{Survey Strategy}

The goal of the NMBS is to obtain high-quality SEDs and redshifts for
galaxies down to a limit of $K\approx 21.5$. In practice, we aim to
secure medium-band photometry with a $1\sigma$ uncertainty of
$\approx 2 \times 10^{-20}$\,ergs\,s$^{-1}$\,cm$^{-2}$\,\AA$^{-1}$.
This constant limit in $F_{\lambda}$ should provide $>8\sigma$
photometry in bands redward of the Balmer break for most
galaxies with $1.5<z<3.5$ and $K<21.5$ (Vega; $K_{\rm AB}<23.3$).
This limit allows us to select and study samples that are $>95$\,\%
complete for galaxies with
stellar masses $>10^{11}$\,\msun\ out to $z\sim 3$
(van Dokkum et al.\ 2006).
The required integration times strongly depend on conditions,
and are typically $\sim 40$\,hrs per band.

We are targeting two fields: a single $27\farcm 6 \times 27\farcm 6$
NEWFIRM pointing within the
COSMOS field ({Scoville} {et~al.} 2007) and a single pointing overlapping
with part of the AEGIS strip ({Davis} {et~al.} 2007). The center of the
COSMOS pointing is at $\alpha = 9^h59^m53.3^s$,
$\delta = +02^{\circ}24^m08^s$ (J2000); it overlaps with the
extensive datasets at other wavelengths that are available for this
field, including the $z$COSMOS deep redshift survey ({Lilly} {et~al.} 2007)
and the upcoming UltraVISTA survey\footnote{http://www.eso.org/sci/observing/policies/PublicSurveys/}.
The AEGIS pointing is centered at $\alpha = 14^h18^m00^s$, $\delta
= +52^{\circ}36^m07^s$ (J2000). It overlaps with about 50\,\% of
the deep ACS and Spitzer imaging in AEGIS and with the
``Westphal'' field, which contains 188
spectroscopically confirmed Lyman break galaxies (see {Steidel} {et~al.} 2003).
For both fields public optical $ugriz$ data are available from
the Deep Canada-France-Hawaii Telescope Legacy Survey\footnote{http://www.cfht.hawaii.edu/Science/CFHTLS/}. These data are of uniform quality (at least
over the extent of the
NEWFIRM fields) and very deep, reaching typical $10\sigma$ AB limits
of $\sim 25.5$ for point sources.

\begin{table*}
\caption{Spectrophotometric Standards\tablenotemark{a}}
\small
\begin{tabular}{lcccccccc}
\hline
\hline
ID & $\alpha$ & $\delta$ & \Ja\ & \Jb\ & \Jc\ & \Ha\ & \Hb\ & $K$ \\
\hline
G191B2B & $05^h05^m30.6^s$ & $+52^{\circ}49^m53.6^s$ & 12.44 & 12.52 & 12.54 & 12.62 & 12.67 & 12.75 \\
GD71    & $05^h52^m27.5^s$ & $+15^{\circ}53^m16.6^s$ & 13.66 & 13.72 & 13.75 & 13.82 & 13.87 & 13.95 \\
GD153   & $12^h57^m02.4^s$ & $+22^{\circ}01^m56.0^s$ & 13.99 & 14.06 & 14.09 & 14.16 & 14.21 & 14.29 \\
P041C   & $14^h51^m57.9^s$ & $+71^{\circ}43^m13.0^s$ & 11.08 & 10.94 & 10.85 & 10.60 & 10.57 & 10.55 \\
P177D   & $15^h59^m13.6^s$ & $+47^{\circ}36^m40.0^s$ & 12.49 & 12.33 & 12.22 & 11.96 & 11.92 & 11.90 \\
P330E   & $16^h31^m33.6^s$ & $+30^{\circ}08^m48.0^s$ & 12.02 & 11.86 & 11.76 & 11.49 & 11.45 & 11.42 \\
1740346 & $17^h40^m34.7^s$ & $+65^{\circ}27^m15.0^s$ & 12.18 & 12.12 & 12.09 & 12.02 & 12.02 & 11.99 \\
1805292 & $18^h05^m29.3^s$ & $+64^{\circ}27^m52.1^s$ & 12.10 & 12.07 & 12.06 & 12.02 & 12.02 & 12.00 \\
1812095 & $18^h12^m09.6^s$ & $+63^{\circ}29^m42.3^s$ & 11.43 & 11.39 & 11.37 & 11.31 & 11.30 & 11.28 \\
KF06T1  & $17^h57^m58.5^s$ & $+66^{\circ}52^m29.3^s$ & 11.89 & 11.84 & 11.61 & 11.31 & 11.30 & 11.29 \\
\hline
\end{tabular}

\tablenotetext{a}{All magnitudes are on the Vega system.}
\end{table*}

These two fields are observed whenever they are available and conditions
are good. In mediocre conditions we observe backup fields to much
shallower depth. The goal is to obtain several of these fields over
the course of the survey, as they play an important role in calibrating
photometric redshifts, constraining
the bright end of the luminosity- and mass-functions, and in providing
targets for follow-up spectroscopy. In the following we report on
observations of the first of these backup fields, the MUSYC
SDSS\,1030+05 field.

\section{Observations and Reduction of the MUSYC SDSS\,1030+05 Field}

The first run of the NMBS comprised a contiguous block of 24 nights,
2008 March 24 -- April 16. A detailed description of the observations
in our primary COSMOS and AEGIS fields will be presented in
K.\ Whitaker et al., in preparation. Here we discuss observations
in the bad weather backup field MUSYC SDSS\,1030+05
(see {Quadri} {et~al.} 2007; {Blanc} {et~al.} 2008). This is one of the
four $30'\times 30'$ fields surveyed by the MUltiwavelength
Survey by Yale-Chile (MUSYC; Gawiser et al.\ 2006). It is centered
on $\alpha = 10^h30^m27.1^s$,
$\delta = +05^{\circ}24^m55^s$ (J2000), and contains the $z=6.3$
QSO SDSSp\,J1030027.10+052455.0 ({Becker} {et~al.} 2001).

We chose this field because it contains
14 objects from a sample of
spectroscopically-confirmed $K$-selected galaxies at $z\sim 2.3$
({Kriek} {et~al.} 2008).
The Kriek et al.\ sample is unbiased with
respect to observed-optical flux, as it was selected on the basis of $K$-magnitude
and photometric redshift only.
Furthermore, rest-frame optical
continuum spectroscopy is available for the entire sample
(see {Kriek} {et~al.} 2008).
The NMBS observations in this field therefore effectively serve
as a small pilot program,
allowing us to assess whether the medium band filter technique
can indeed provide reliable SEDs and
redshifts for optically-faint sources.
Even though there are thousands of spectroscopic redshifts available
in our primary survey fields
(AEGIS and COSMOS), they are almost exclusively for optically-bright,
low redshift galaxies.

Because the SDSS\,1030+05 field served as a bad weather backup field the
data quality is relatively poor (as compared to the main
survey fields). The seeing was typically in the range $1\farcs 4-1\farcs 8$
when we observed the field, and the background was often relatively
high. Total integration times were 2.3\,hrs in \Ja,
1.1\,hrs in \Jb, 2.5\,hrs in \Jc, 1.1\,hrs in \Ha, and 1.6\,hrs in \Hb.
The data were processed using a new reduction package, which was developed by
one of us (IL). The heart of the code
is similar to the popular IRAF {\sc xdimsum}
package, but it incorporates many of the changes that we have developed over
the years (see {Labb{\' e}} {et~al.} 2003; {Quadri} {et~al.} 2007). The code was completely
rewritten in the IDL programming environment,
automated, and optimized for NEWFIRM.

\begin{figure*}[t]
\epsfxsize=13.5cm
\epsffile[74 462 351 690]{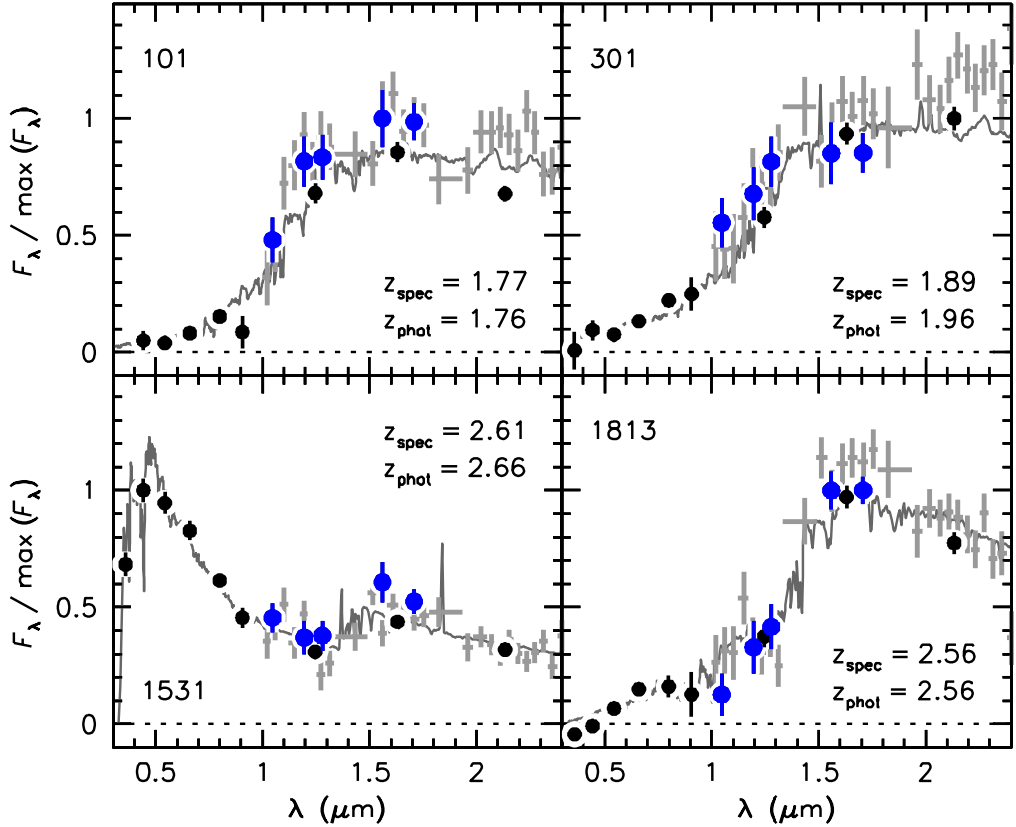}
\caption{\small Spectral energy distributions from $0.3 - 2.4\,\mu$m
of the four galaxies in the SDSS\,1030 Kriek et al.\ (2008) sample
with the highest S/N ratio. Black points are broad band photometric
data, blue points are the new medium band data.
The medium band
data are able to pinpoint the location of rest-frame optical breaks
in the spectra.  Dark grey spectra are the best-fit EAZY model SEDs.
Light grey points are
binned near-IR spectra obtained with GNIRS on Gemini, from Kriek
et al. The best-fit model SEDs fit the (independent!) GNIRS spectra
very well.
\label{seds.plot}}
\end{figure*}

The data were calibrated by repeated observations of six different near-IR
spectrophotometric standards. Synthetic magnitudes of these stars
were calculated by integrating their observed (HST/NICMOS) spectra
in our filters. We verified that this method reproduces the (independently
calibrated) broad-band $J$, $H$, and $K$ band magnitudes of these stars.
Based on the observed variation of zeropoints derived from
different stars and on different photometric nights we estimate that
the zeropoint uncertainties are $\lesssim 0.02$ mag.
The stars and their synthetic magnitudes are listed in Table 2, along
with stars that were not observed during the March/April run.
The achieved $5\sigma$ depths in the SDSS\,1030+05 field are
$J_1 = 22.2$, $J_2 = 21.5$, $J_3=21.4$, $H_1=20.8$, and $H_2=20.9$
(total Vega magnitudes for point sources).

The medium-band images were combined with $UBVRIzK$
imaging from the MUSYC survey. A general description of the MUSYC optical
imaging is given in {Gawiser} {et~al.} (2006); specific aspects of the SDSS\,1030
field are provided in {Quadri} {et~al.} (2007) and
{Blanc} {et~al.} (2008). The methodology
for PSF-matching of the optical-
and near-IR data and the procedures for creating a
$K$-selected catalog are the same
as used in {Labb{\' e}} {et~al.} (2003) and {Quadri} {et~al.} (2007). We also included
Spitzer IRAC imaging in the analysis, which are
described in Marchesini et al.\ (2008). The IRAC data
do not cover the full NEWFIRM field, but do cover the galaxies
from the {Kriek} {et~al.} (2008) sample.
The final product is a $K$-selected
catalog with accurate $UBVRIzJ_1J_2J_3H_1H_2K$+IRAC
fluxes.

\section{Results}

\subsection{Spectral Energy Distributions}

Of the 14 galaxies in the SDSS\,1030 field that overlap with the
{Kriek} {et~al.} (2008) sample, four have an average S/N ratio in the
\Ha\ and \Hb\ filters which exceeds our
survey criterion of 8. The broad + medium band
photometry of these galaxies is shown in Fig.\ \ref{seds.plot},
along with the (binned) GNIRS spectra from {Kriek} {et~al.} (2008)
(shown in light grey).

The medium band photometry is consistent with both the GNIRS
continuum spectroscopy and with the
broad band $J$ and $H$ data. The medium bands
sample the spectral energy
distributions (SED) more finely than the broad bands,
and capture the overall shape of the SEDs as traced by the GNIRS
spectra. In particular, there are obvious breaks in
the medium band photometry, making it possible to pinpoint the
location of the redshifted Balmer- or 4000\,\AA\ break {\em within}
the $J$ band window (or between \Jc\ and \Ha\
for objects 1531 and 1813).
This is a significant advance: the medium bands sample
the SED with a spectral resolution of $R=10-11$, whereas
standard broad-band near-IR photometry corresponds to $R=3-4$.

The ability to detect breaks obviously depends on a combination
of the intrinsic strength of the break and the S/N ratio of the
photometry. Galaxy 1813\footnote{The numbering follows {Kriek} {et~al.} (2008),
who give coordinates, $K$ magnitudes, and other information for
these galaxies.} has a very strong break between \Jc\ and \Ha,
which would also have been detected in shallower exposures.
The breaks in the other galaxies are weaker, and the ability to
measure accurate redshifts will depend sensitively
on the quality of the photometry. The average S/N
ratio in \Ha\ and \Hb\ ranges from 8 to 15 for the
four galaxies shown in Fig.\ \ref{seds.plot};
the goal of the main survey is to reach a S/N ratio $>8$
in the bands redwards of $\lambda_{\rm rest}=4000$\,\AA\ for
galaxies with $K<21.5$ in our NEWFIRM fields.

\subsection{Photometric Redshifts}

Redshifts are measured with the photometric redshift code EAZY
(Brammer et al.\ 2008), which is optimized for situations
where complete spectroscopic calibration samples are not available.
The default template set and rest-frame template error function
were used, and the default magnitude and redshift priors (appropriate
for the $K$ band).
Although not of great consequence
in the present context, the ``CHI2\_SCALE'' parameter was
set to 0.5 in order to provide more realistic errorbars.\footnote{We
find that the default EAZY uncertainties slightly overestimate the errors;
as discussed in Brammer et al.\ (2008) the exact interpretation
of the uncertainties can vary between datasets.}

\begin{figure}[h]
\epsfxsize=7cm
\epsffile{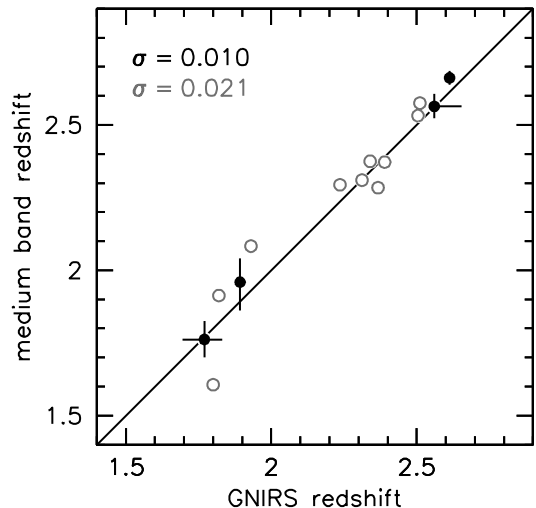}
\caption{\small Comparison of photometric redshifts derived from
medium band photometry to spectroscopic redshifts measured with
the GNIRS near-IR spectrograph on Gemini for the four galaxies
shown in Fig.\ 2 (solid symbols).
There is very good agreement, with scatter $0.01-0.02$
in $\Delta z/(1+z)$. Open symbols show the remaining 10 objects
from the Kriek et al.\ (2008) sample. The scatter is
small even for these galaxies, even though the S/N of their
medium band photometry is lower than our survey criterion.
\label{photzspecz.plot}}
\end{figure}

The photometric redshifts are compared to the Gemini/GNIRS
redshifts from {Kriek} {et~al.} (2008) in Fig.\ \ref{photzspecz.plot}.
There is excellent agreement for the four galaxies, with
the biweight scatter in $(z_{\rm phot}-z_{\rm spec})/(1+z_{\rm spec})$
only 0.010. This value is not very robust given the small sample
size (the normalized median absolute deviation is 0.020, and the
rms is 0.011), but it is substantially better than what has so far been
achieved at these redshifts.
As an example, Grazian et al.\ (2007) and Brammer et al.\ (2008) 
find a scatter of 0.06 -- 0.07 at $z>1.5$ using state of the art broad-band data
in the CDF-South field. Similarly, Ilbert et al.\ (2008) find a scatter
of 0.06 (with 20\,\% catastrophic outliers) at $1.5<z<3$ in COSMOS
using 30 photometric bands (including GALEX, IRAC, and
18 medium-bandwidth optical filters from Subaru).

Open symbols in Fig.\ \ref{photzspecz.plot} have a S/N less than our
survey criterion of 8 in the \Ha\ and \Hb\ bands. Interestingly, they
nevertheless show relatively small scatter as well, and the
scatter in the full sample of 14 galaxies is $\approx 0.023$
in $\Delta z / (1+z)$. The sample is obviously too small to investigate
the exact behavior of the redshift uncertainty as a function of
S/N ratio, but the presently available information suggests that
uncertainties of $\approx 0.02$ in $\Delta z / (1+z)$ can be
achieved with a S/N of 6--10 in filters redward of the
redshifted Balmer/4000\,\AA\ break.

The dark grey spectra in Fig.\ \ref{seds.plot} are the best-fitting
EAZY templates.\footnote{Each template is actually a linear combination of six
``base'' templates, and these six base templates are themselves
linear combinations
extracted from a large template library -- see Brammer et al.\ (2008)
for details.} These templates fit the observed photometry well,
and all four galaxies have acceptable best-fit $\chi^2$ values
(partly owing to the template error function; see Brammer et al.\ 2008).
The dark grey model spectra fit the binned GNIRS spectra remarkably well;
the only significant deviation is a $\approx 20$\,\%
underprediction of the flux at $\approx 2.2\,\mu$m for galaxy 301.

%
%

\section{Other Applications: Selecting the Coolest Brown Dwarfs}
\label{tdwarf.sec}

The medium-band
filters were designed to improve redshift estimates and stellar
population constraints for 
distant galaxies but can also be used for other purposes, in
particular when used in a wide, relatively shallow survey.
Among these other applications are the identification of
objects with extremely bright emission lines; improved
star/galaxy separation; identification and characterization of
high redshift galaxies and QSOs; and finding cool brown dwarfs.
In the following we expand on the latter application.

\begin{figure}[h]
\epsfxsize=7cm
\epsffile{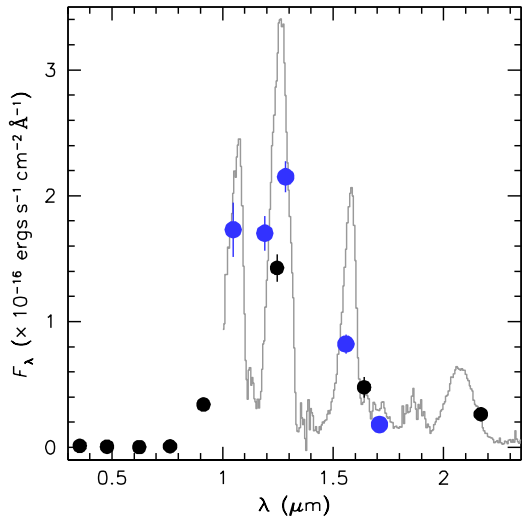}
\caption{\small
Observations in $ugrizJHK$ (black) and the medium band filters
(blue; obtained in twilight)
for the T7 dwarf 2MASS\,1553+1532. The spectrum is from
Burgasser et al.\ (2002). This cool dwarf has a unique signature in the medium
band filters, particularly in the $H_1-H_2$ color.
\label{tdwarf.plot}}
\end{figure}

As illustrated for a T7 dwarf in Fig.\ \ref{tdwarf.plot}
the spectra of very late type dwarfs (beyond the L/T boundary)
are characterized by strong H$_2$O
and CH$_4$ absorption. The subtype within the T class is determined
by the strengths of these absorption bands, which in turn are thought to
be closely correlated with the effective temperature
(see {Burgasser} {et~al.} 2002). The
most dramatic change going from T1--T9 is in the broad
methane absorption at $\sim 1.7\,\mu$m, which is weak
at the L/T boundary and almost complete for the coolest T dwarfs.

Finding the coolest dwarfs typically involves a multi-stage process:
the initial selection uses $JHK$ photometry
from 2MASS (e.g., {Burgasser} {et~al.} 2002) or $iz$ photometry from
the CFHTLS (e.g., {Delorme} {et~al.} 2008); follow-up broad-band near-IR
imaging is used to weed out interlopers and spurious sources and to obtain
accurate $J-H$ and $H-K$ colors; and near-IR spectroscopy
provides the spectral type.

\begin{figure}[h]
\epsfxsize=7cm
\epsffile{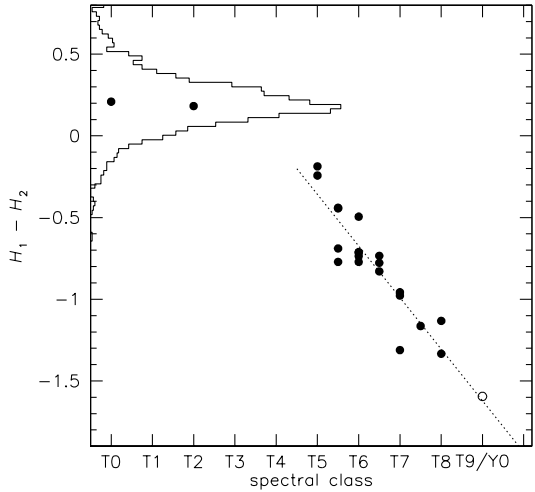}
\caption{\small Relation between $H_1-H_2$ color and T subclass, as
derived from near-IR spectra of cool dwarfs from Burgasser et al.\ (2002).
The histogram shows the distribution of $H_1-H_2$ colors for all objects
in the SDSS\,1030 field; dwarfs with spectral type
$\gtrsim$ T5 can be uniquely identified by their extremely
blue $H_1-H_2$ color.
Also included is the coolest dwarf known to date, which may be a
T/Y boundary object (Delorme et al.\ 2008). This object falls on
the same relation as the other T dwarfs.
\label{diagnostic.plot}}
\end{figure}

Interestingly, the medium band filters offer an extremely efficient way
to select ultra-cool stars, as the \Hb\ filter coincides almost exactly
with the
location and width of the CH$_4$ feature at $\sim 1.7\,\mu$m.
As a result, the $H_1-H_2$
color is a very strong function of spectral class (and hence
effective temperature) for the coolest dwarfs. In Fig.\ \ref{diagnostic.plot}
we show the relation
between the $H_1-H_2$ color and spectral type for the T dwarfs of
{Burgasser} {et~al.} (2002). The colors are not based on models but
were calculated by integrating the
observed near-IR spectra of these dwarfs\footnote{Obtained from
http://web.mit.edu/ajb/www/tdwarf/\#spectra.}. 

There is a clear relation between T subclass and $H_1-H_2$ color,
with $H_1-H_2$ progressively bluer for
later types. The relation shown by the broken line
has the form T$x = 4.2 - 2.7 (H_1-H_2)$; it has an rms of less than
one subclass. The histogram shows the distribution of $H_1-H_2$ colors
for all objects in the SDSS\,1030 field. Normal stars and galaxies
are well-separated from the coolest dwarfs, and
Fig.\ \ref{diagnostic.plot} suggests that a simple
selection on $H_1-H_2$ color (e.g., $H_1-H_2<-1$)
should yield a clean sample of very cool dwarfs.

It is interesting to speculate whether the $H_1-H_2$ color could
also be used to select stars with $T_{\rm eff}<700$\,K,
the elusive ``Y'' class (e.g., {Kirkpatrick} {et~al.} 1999).
The CH$_4$ band saturates near the T/Y boundary, which limits
its utility for spectral classification. Nevertheless, the recently
discovered T/Y transition object CFBDS\,J005910.90--011401.3
({Delorme} {et~al.} 2008) falls on the same relation as the late T dwarfs
(see Fig.\ \ref{diagnostic.plot}). It may therefore be possible
to select Y dwarfs by the simple criterion $H_1-H_2<-1.5$.

\section{Conclusions}

We have developed a medium bandwidth filter system in the near-IR, providing
a compromise between spectroscopy and broad-band imaging. Installed in
the wide-field NEWFIRM camera on Kitt Peak, the filters enable us to
obtain high quality redshifts and
spectral energy distributions for large, complete samples of galaxies
with far greater efficiency than is possible with spectroscopy.
The NMBS aims to obtain redshifts of $\approx 40,000$
galaxies with $K< 21.5$, some 8000 of which are expected to be at $z>1.5$.
To put this in context,
with a multi-object near-IR spectrograph such as FLAMINGOS-2 on Gemini
it would require $\sim 2500$\,hrs to obtain redshifts for 1000 galaxies to
this limit (scaling from Kriek et al.\ 2008).

Although the initial results reported here are promising,
the accuracy of the redshift measurements needs to be verified.
The medium-band technique relies on the presence of a
break in the rest-frame optical, and the improvement in
photometric redshift estimates will therefore depend on galaxy type.
Very young stellar populations with ages $\lesssim 300$\,Myr do not
have a significant Balmer break, and the accuracy of the redshifts
of many Lyman break and ``BM/BX'' galaxies (Steidel et al.\ 2004)
may therefore not be much better than can be derived from
broad-band optical photometry alone. Similarly, very dusty galaxies
can have featureless red SEDs.

With larger samples of galaxies
with spectroscopic redshifts we will be able to quantify these and
other effects (such as the presence of bright emission lines, and
the redshift-dependence of redshift errors). Such
spectroscopic samples will obviously not be representative of
our entire sample, but they can be used to
assess the reliability of the uncertainties given by the EAZY code.
If accuracies of 0.01 -- 0.02 turn out to be typical for galaxies
down to our survey limit, the NMBS will establish the relations
between redshift, color, and density  at $1.5<z<3.5$ with excellent
statistics. Reduced images, catalogs, and derived redshifts, stellar
population parameters, and rest-frame colors will be publicly released
after the survey is completed. Finally, we note that the
\Ha\ and \Hb\ filters enable very efficient searches for
late T and candidate Y dwarfs.

\acknowledgments

We thank Ron Probst and the staff at NOAO for the construction of
NEWFIRM and for their extensive help with, and enthusiasm for,
this project. Philippe Delorme and Linhua Jiang kindly made spectra
available to us in digital form. Support from NSF CAREER grant
AST-0449678 and from NSF grant AST-0807974 is gratefully acknowledged.
We thank the anonymous referee for constructive comments which
improved the manuscript.


\clearpage

\end{document}